\documentclass[useAMS,usenatbib]{mn2e}

\usepackage{epsfig}

\def\lesssim{\mathrel{\hbox{\rlap{\hbox{\lower4pt\hbox{$\sim$}}}\hbox{$<$}}}}
\def\gtrsim{\mathrel{\hbox{\rlap{\hbox{\lower4pt\hbox{$\sim$}}}\hbox{$>$}}}}

\title[Eccentric particles near an eccentric planet]
{Chaotic zone boundary for low 
free eccentricity particles near an eccentric planet}

\author[Quillen \& Faber]
{Alice C. Quillen  \& Peter Faber  \\
Department of Physics and Astronomy, University of Rochester, 
Rochester, NY 14627; 
aquillen@pas.rochester.edu; pfaber@mail.rochester.edu}

\begin{document}

\label{firstpage}

\maketitle

\begin{abstract}
We consider particles with low free or proper eccentricity
that are orbiting near planets on eccentric orbits.  
Via collisionless particle integration
we numerically find the location of the boundary 
of the chaotic zone in the planet's corotation region.
We find that the distance in semi-major axis between the planet and boundary 
depends on the planet mass to the 2/7 power and is independent of
the planet eccentricity, at least for planet eccentricities below 0.3.  
Our integrations reveal a similarity between the dynamics of particles
at zero eccentricity near a planet in a circular orbit and 
with zero free eccentricity particles near an eccentric planet.
The 2/7 law has been previously explained by estimating
the semi-major at which the first order mean motion resonances 
are large enough to overlap.
Orbital dynamics near an eccentric planet could differ due to first order
corotation
resonances that have strength proportional to the planet's eccentricity.
However, we find the corotation resonance width at low free
eccentricity is small.  Also the first order resonance width 
at zero free eccentricity is
the same as that for a zero eccentricity particle near a planet
in a circular orbit. 
This accounts for insensitivity of the chaotic zone width
to planet eccentricity.
Particles at zero free eccentricity near an eccentric planet
have similar dynamics
to those at  zero eccentricity near a planet in a circular orbit.

\end{abstract}

\begin{keywords}
celestial mechanics; 
planetary systems : protoplanetary disks
\end{keywords}

\section{Introduction}

Chaotic diffusion associated with the overlap of resonances has
been shown to be responsible for instabilities in the solar system
(e.g., see \citealt{holman93,lecar01,lecar02,tsiganis05}).
For the restricted 3-body problem
\citet{wisdom80} first showed that the width of the chaotic zone near a planet
could be explained by calculating the location at which
the first order mean motion resonances are large enough to overlap.
The zone width has been measured numerical and predicted theoretically
\citep{wisdom80,duncan89,murray97} for a planet in a 
circular orbit, though 
some work has extended the stability
analysis for bodies in orbits near circular and eccentric binary stars
\citep{holman99,mudryk06}.
The stability of bodies at low eccentricity 
residing in multiple planet extrasolar systems 
have also been investigated numerically
(e.g., \citealt{rasio96,lepage04,barnes04}).  

Recently \citet{quillen06b} suggested 
that the edge of Fomalhaut's eccentric ring could
be due to truncation by a 0.1 eccentricity Neptune mass planet.  
The nearby star Fomalhaut
hosts a ring of circumstellar material \citep{aumann85,gillett85}
residing between 120 and 160 AU from the star
\citep{holland98,dent00,holland03}.
{\it Spitzer Space Telescope} infrared
observations of Fomalhaut reveal a strong brightness asymmetry
in the ring \citep{stapelfeldt04,marsh05}.  
Recent {\it Hubble Space Telescope} ({\it HST}) observations show
that this ring has both a steep
and eccentric inner edge \citep{kalas05}.
%
The sharp disk edge suggested that the dust particles
are in orbits with low free or proper eccentricity, thus
the ring has eccentricity equal to the forced eccentricity
caused by secular perturbations from the proposed planet.
Such a configuration  is possible if inelastic collisions in the disk
damp the eccentricities of particles, resulting in a particle
distribution that moves along nearly closed streamlines
or closed and non-self-intersecting orbits.   
Fomalhaut's ring has a intermediate
collision timescale of $10^3$ orbits, 
estimated from its normal disk opacity $\tau \sim 1.6\mu$m
at 24$\mu$m \citep{marsh05}.

While previous theoretical and numerical studies have 
considered orbit stability near  the corotation region for
planets on circular orbits, little work has been done
considering the stability of orbits near
a planet on an eccentric orbit.  
The dynamical problem of an object orbiting a planet in a circular orbit
has a conserved quantity, the Jacobi integral, that is not conserved
when the planet is on an eccentric orbit.
Surfaces of section have
been used to illustrate the types of orbits (tori or area filling)
for the restricted three-body system \citep{wisdom85,winter97}.
However, when the planet is eccentric there is no
extra integral of motion making it difficult to create surfaces of section.   
We are motivated here to consider the role of the 
planet's eccentricity in setting the boundary of non-stochastic orbits
in the corotation region.  We focus here on particle orbits
that have nearly zero free eccentricity and so have
eccentricity set by the forced eccentricity due to secular
perturbations from the planet.

\section{Numerical integrations}

Numerical integrations were carried out in the plane,
using massless and collisionless particles under the gravitational
influence of only the star and a
massive planet with eccentricity, $e_p$, using a
 conventional Burlisch-Stoer numerical scheme.
A particle near a planet on an eccentric orbit feels  secular
perturbations from the planet   if it is located away from low-order
mean motion resonances. 
The particle's eccentricity and longitude
of periastron precess about a point set 
by the distance to the planet, the planet's eccentricity 
longitude of periastron.  
The secular motion can be described in terms of a free (or proper)
and forced eccentricity (e.g., \citealt{M+D}).
Only a particle with zero free eccentricity would have a fixed argument 
of periastron and eccentricity.
In our integrations 
the initial particle eccentricities were set to
the predicted forced eccentricity due to secular
perturbations from the planet and the
and longitudes of periastron were chosen to be identical to that of
the planet.
Initial mean anomalies were randomly chosen.
Particles were
removed from the integration when their eccentricity was larger than 
$e_{max}=0.9$. 
We work in units of the planet's semi-major axis, $a_p$, and orbital period.
The mass of the planet is described
in terms of its mass ratio, $\mu$, the ratio of the planet
mass to that of the central star. 

\subsection{Measurement of the width of the chaotic zone }

\begin{figure}
\includegraphics[angle=0,width=3.6in]{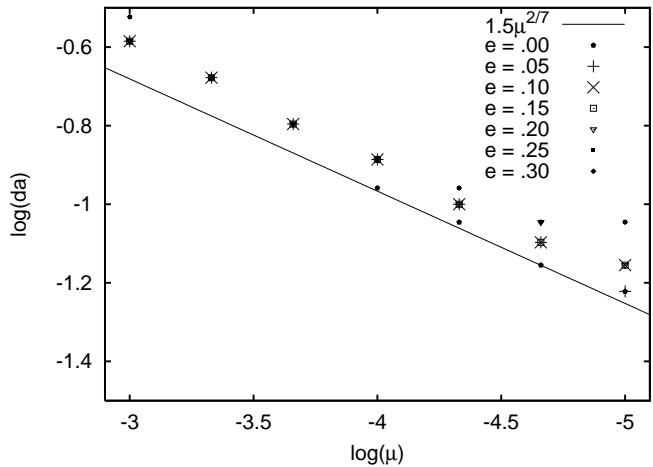}
\caption{
The distance between the planet semi-major axis
and that of the chaotic zone boundary is plotted as a function
of planet mass ratio, $\mu$.
Points are plotted at the semi-major axis  at which particles
lifetime exceeded $10^4$ orbits.
Each point type corresponds to different planet eccentricity.
The points lie on top of one another because the chaotic zone width
is independent of the planet eccentricity for orbits with
zero free eccentricity.
Particles were initially placed in orbits with zero free eccentricity and
arguments of peri-astron the same as the planet.
The solid line corresponds to that predicted by equation \ref{eqn:27}.
\label{fig:da}
}
\end{figure}

As a function of initial semi-major axis we measured
the lifetimes of particles before removal from the integration.  
The semi-major axis bins for each lifetime measurement 
had width $da=0.01$,  and 100 particles were integrated for each bin.
An abrupt increase
in the particle lifetime was seen in the simulations as a function
of initial semi-major axis (see Figure 2 by \citealt{quillen06b}).  
Inside the chaotic zone particles
were scattered to high eccentricity and removed 
from the simulation, but outside it,
the lifetime of particles
exceeded $10^4$ orbits.   For planet masses ratio
between $10^{-3}$ and  $10^{-5}$ and eccentricities between
$0.05$ and $0.2$ we measured the semi-major
axis, $a_z$, at which this transition in lifetime occurred. 
The distance $da  = {a_z - a_p \over a_p}$ between this
semi-major axis and the planet's divided by the planet's semi-major
axis is plotted in Figure \ref{fig:da}.  

In Figure \ref{fig:da} we note that the chaotic zone boundaries
depend on the planet mass. The scaling is consistent with
that predicted from the 2/7 law or 
\begin{equation}
da = 1.5 \mu^{2/7}
\label{eqn:27}
\end{equation}
\citep{wisdom80} where the constant 1.5 is taken from numerical measurements
by \citet{duncan89}.
The  offset between the line predicted by equation \ref{eqn:27} is
not significant as we have measured the width from an ensemble of particles
and required them all to remain at least $10^4$ orbits. This implies that 
we have measured a location outside 
the last stable orbit in a surface section
or a closed orbit radius vs energy bifurcation plot.
The points shown in Figure \ref{fig:da} include those integrated
for a zero eccentricity planet.
so the offset
between that predicted by the 2/7th law is the same for the higher 
planet eccentricities as for the zero planet eccentricity.
Using finer spacing in semi-major axis and by restricting the 
initial orbital elements rather than choosing them randomly,
it is possible to find stable orbits somewhat closer to the planet.
The offset from the predicted 
line is caused by the measurement procedure, rather than
the planet eccentricity.  
We note an increased scatter in the
chaotic zone boundary at lower masses in Figure \ref{fig:da}, however we
find no clear trend in the scatter as a function of planet eccentricity.
Since the particle lifetimes in the chaotic zone 
are longer for the lower mass planets 
more particle trajectories may need to be integrated
to achieve the same precision in the measurement of the chaotic zone
boundary at lower planet masses.
The choice of initial random mean anomalies 
could also contribute to the scatter at lower planet masses.

Figure \ref{fig:da} shows points corresponding
to integrations with different planet eccentricities.
The points for integrations near
eccentric planets lie on top of those at low or zero
planet eccentricity.  In other words the width of the chaotic
zone appears to be independent of the planet eccentricity.
We find that there exists a long lived 
low free eccentricity region near moderately eccentric
planets (in the planar problem).  
This region has $da < e_p$ for the more highly eccentric planets
and so is nearer the planet's major axis than the planet's periastron.
In other words,
these orbits pass closer to the star than 
the periastron of the planet, but because they are apsidally aligned with
the orbit of the planet they do not cross the planet's orbit.

A disk with a low collision rate, could evolve to a distribution with
nearly closed orbits fixed about the forced eccentricity and apsidally
aligned with a planet.  Our planar numerical integrations
show that this region is stable over long timescales 
(greater than $10^4$ orbits)
and so could host a long lived  planetesimal distribution.

\subsection{Dispersion and lifetimes}

\begin{figure}
\includegraphics[angle=0,width=3.5in]{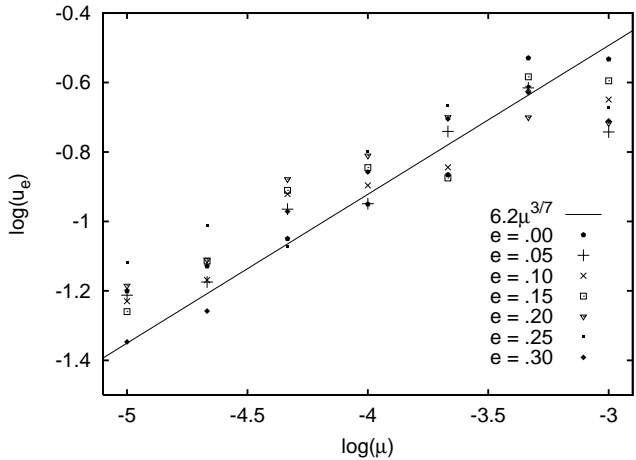}
\caption{
Eccentricity dispersion, $u_e$, 
in the disk edge as a function of planet mass.  The line shown is
$u_e =  6.2\mu^{3/7}$ and consistent with that predicted with
equation \ref{eqn:ue} for particles near a planet in a circular orbit.
Each point type corresponds to integrations containing a planet with
a different eccentricity.
The eccentricity dispersion is
not strongly dependent on the planet eccentricity.
\label{fig:ue}
}
\end{figure}

We have also  measured
the eccentricity distribution 
after $10^4$ planetary orbits in the disk edge. 
These are shown in Figure \ref{fig:ue}.  
Each point shown
on this figure correspond to measurements
based on integration of 100 particles.
The corotation chaotic
region arises from resonance overlap. However outside the chaotic zone
mean motion resonances exist that affect the particles, though because
they do not overlap other mean motion resonances the particles 
do not vary their orbital parameters stochastically or do vary stochastically
but on  much longer timescales.
Figure \ref{fig:ue} shows that the velocity dispersion in the stable
boundary is likely to depend on planet mass and not eccentricity.

\begin{figure}
\includegraphics[angle=0,width=3.5in]{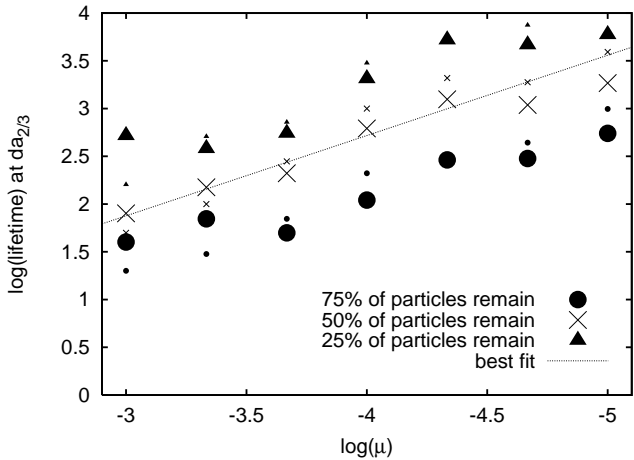}
\caption{
Particle lifetime in planetary orbits as a function of planet mass for particles
with initial semi-major axis 2/3 of the way between the planet's
semi-major axis and the chaotic zone boundary.
Large and small points are for planet eccentricity $e_p = 0.05$
and $0.2$, respectively.
The circles, crosses and triangles  correspond to times when 75\%, 50\% and
25\% of the particles remain in the simulation, respectively.
The line shown has lifetime equal to $ 0.23 \mu^{-0.84}$. 
Particles have initial angle of pericenter identical to that
of the planet and zero free eccentricity.
The lifetimes are not strongly dependent on the planet eccentricity.
\label{fig:life}
}
\end{figure}

To characterize the lifetime of particles in the chaotic zone
we consider particles within initial semi-major axis at
2/3 the distance in semi-major axis to the zone edge.
The timescale for removal of 25\%, 50\% and 75\% of the particles
is plotted as a function planet mass for planet eccentricities, $e_p =0.05$
and 0.2 in Figure \ref{fig:life}.
Again particles are initial started in orbits with zero free eccentricity.
From comparing Figure \ref{fig:life}, we see no significant difference 
in the particle lifetimes  at the two planet eccentricities.
Particles placed into the chaotic zone at
zero free eccentricity, would have the a similar resident lifetime as 
those placed at zero eccentricity near a planet of zero eccentricity.

For orbits with eccentricity equal to the forced eccentricity we find 
that there is a long lived or stable
region in close proximity to eccentric planets.  Inside this region
particles are pumped to  high eccentricity and scattered by the planet
on a timescale orders of magnitude faster than outside this region.
We have found that the width of this chaotic zone, 
in semi-major axis is independent of the planet eccentricity. 
The eccentricity dispersion in the disk edge and
the lifetime of particles within the zone is also nearly independent
of planet eccentricity.  These results suggest there is a similarity
in the dynamics of particles at zero free eccentricity
and those at zero eccentricity
near a planet in a circular orbit.  In the following section
we explore a Hamiltonian formulation that shows that this is in 
fact the case.

\section{Hamiltonian Formulation}

In this section we reconsider the theory of mean motion resonance overlap.
We consider the role of first order secular perturbations from
the planet and the two angular perturbations that are
associated with each first order mean motion resonance.
We follow the notation by \citet{quillen06a}.
We employ the Poincar\'e coordinates
\begin{displaymath}
\lambda = M + \varpi, \qquad \gamma = -\varpi
\end{displaymath}
and their associated momenta
\begin{displaymath}
L = \sqrt{GM_* a}, \qquad \Gamma  = \sqrt{GM_* a}(1-\sqrt{1-e^2})
\end{displaymath}
where $M_*$ is the mass of the star,
$\lambda$ is the mean longitude, $M$ is the mean anomaly, $\varpi$
is the longitude of pericenter, $a$ is the semimajor axis, and $e$ is
the eccentricity. These variables are those
describing the orbit of a particle or planetesimal in a plane.
The Hamiltonian for the Keplerian system in these coordinates restricted
to a plane is
\begin{displaymath}
H(L,\lambda;\Gamma, \gamma) = -{GM_* \over 2 L^2} - R
\end{displaymath}
where $R$ is the disturbing function, proportional to the planet
mass, that depends on the
coordinates of the particle and on the coordinates
of the planet.  The planet's semimajor axis and mass are denoted $a_p$
and $m_p$, respectively. The planet's other coordinates
are subscripted in the same way.
The mean motion of the particle $n = \dot \lambda$ where
$\dot \lambda$ is the derivative with respect to time of $\lambda$.

Hereafter we adopt a unit convention with distances
in units of the planet's semi-major axis, $a_p$.
Time is put in units of $\sqrt{a_p^3/GM_*}$.
We define $\mu$ to be the mass ratio $\mu \equiv m_p /M_*$.
At low eccentricity, $\Gamma/L \approx e^2/2$, relating the momentum $\Gamma$
to the particle eccentricity.
We often give the particle semimajor axis in terms of
the variable $\alpha \equiv {a_p \over a}$ if $a>a_p$
(external to the planet)
and $\alpha \equiv {a \over a_p}$ for $a<a_p$ (internal to the planet).

The unperturbed Hamiltonian or that lacking the disturbing function
\begin{displaymath}
H_0(L,\lambda; \Gamma, \gamma)  = -{1\over 2 L^2}
\end{displaymath}
We consider the the $j:j-k$ exterior mean motion resonance (planet
is an interior perturber).
We perform a canonical transformation using the mixed variable
generating function
\begin{displaymath}
F_2 = I (j \lambda - (j-k)\lambda_p)
\end{displaymath}
where $\lambda_p = n_p t$,
leading to new variables\footnote{There is an 
error in the $I$ variable definition by \citep{quillen06a}.}
\begin{displaymath}
I =   L/j, \qquad  \psi =  j\lambda - (j-k) \lambda_p
\label{eqn:ILj}
\end{displaymath}
and new Hamiltonian
\begin{displaymath}
H_0'(I , \psi; \Gamma, \gamma)
  =  {- 1 \over 2 j^2  I^2} -  {(j -k) } I n_p.
\end{displaymath}

We now expand around a particular value for $I$. 
Let
\begin{equation}
\Lambda \equiv  I - I_0.
\label{eqnI0}
\end{equation}
In terms of the particles mean motion
$I_0 = n^{-1/3}/j$.
Since we have adopted units $n_p = 1$, we find
$I_0 = \alpha^{-1/2}/j$ where
$\alpha = {a_p \over a}$.
Our Hamiltonian now reads
\begin{displaymath}
K_0(\Lambda, \psi; \Gamma, \gamma) =
   {\rm constant \ } + (jn -  (j -k)n_p)\Lambda
 - {3 \Lambda^2 \over 2 j^2 I_0^4}. \nonumber
\end{displaymath}
We can write the unperturbed Hamiltonian as
\begin{displaymath}
K_0(\Lambda, \psi; \Gamma, \gamma)
  = a' \Lambda^2 +  b'\Lambda  + {\rm constant}
\end{displaymath}
with coefficients
\begin{eqnarray}
a' & = & - {3 \over 2} j^2 \alpha^2 \\
b' & = & nj - (j-k)n_p = \alpha^{3/2}j - (j-k) n_p, \nonumber
\label{eqn:coefs}
\end{eqnarray}
similar to Equation 2 by \citet{quillen06a}.
Exactly on resonance $\alpha = \left({j-k\over j}\right)^{2/3}$
and $b' =0$.
The primes here are given to differentiate $a'$ from
the semi-major axis $a$.

We now recover the disturbing function that is
is traditionally expanded as a cosine series of angles in orders
of planet and particle eccentricity.
We keep the terms inducing precession of the longitude
of periapse and first order (in eccentricity)  
or $k=1$ terms containing $\psi$
and $\varpi$.
The full Hamiltonian
\begin{eqnarray}
K(\Lambda, \psi; \Gamma,\gamma)
   &  = &
   a' \Lambda^2 +   b' \Lambda +
   c' \Gamma 
  \qquad
\label{eqn:hamgen}
\\ 
& &  
 +  d' \Gamma^{1/2} \Gamma_p^{1/2} \cos(\varpi - \varpi_p)
\nonumber
\\
& &  
  + g_0 \Gamma^{1/2}   \cos{( \psi -  \varpi })
\nonumber
\\
& &  
  + g_1 \Gamma_p^{1/2} \cos{( \psi -  \varpi_p })
\nonumber
\end{eqnarray}
where $\Gamma_p \equiv {e_p^2 L \over 2 }$ and
with the following coefficients \footnote{
Eqn 4 by \citet{quillen06a} for $c$ should have had a factor of $\alpha^{3/2}$.}
\begin{eqnarray}
c'  &=&  - {\mu \over 4} \alpha^{5/2} b^1_{3/2} 
\nonumber \\
d'  &=&    {\mu \over 2 } \alpha^{5/2} b^2_{3/2} 
\nonumber \\
g_0 &=&  - \mu \sqrt{2} \alpha^{5/4} f_{31} 
\nonumber \\
g_1 &=&  - \mu \sqrt{2} \alpha^{5/4} f_{27}.
\label{eqn:gcofs}
\end{eqnarray}
The coefficients $f_{31}$ and $f_{27}$ are given in Table B.7 by \citep{M+D}.
The functions  $b^1_{3/2}$ and $b^2_{3/2}$ are Laplace coefficients
and are functions of $\alpha$.
The approximate asymptotic limits for large $j$ and $\alpha \to 1$ are
$f_{31} \to j$ and $f_{27} \to -j$.
We have used the approximation $e^2 \sim 2 \Gamma \alpha^{1/2}$.
The term proportional to $\cos(\psi - \varpi)$ is often
called the $e$-resonance since $\Gamma^{1/2} \propto e$.
The other term can be called an $e'$-resonance
or a corotation resonance since it
does not depend on the particle's longitude of perihelion or $\varpi$.

We first consider secular perturbations only,
ignoring the $g_0,g_1$ terms and considering the following
\begin{equation}
K(\Gamma, \gamma) = c' \Gamma  
 +  d' \Gamma^{1/2} \Gamma_p^{1/2} \cos(\varpi - \varpi_p).
\end{equation}
We find a fixed point at
\begin{eqnarray}
-\gamma &=& \varpi  =  \varpi_p \nonumber \\
\Gamma^{1/2}_{f} &=& {b^2_{3/2} \over b^1_{3/2}} \Gamma_p^{1/2}
\end{eqnarray}
where $\Gamma_f \approx e_{forced}^2 L /2$.  This fixed point is
equivalent to a closed orbit with eccentricity equal
to the forced eccentricity 
\begin{displaymath}
e_{forced} = {b^2_{3/2} \over b^1_{3/2}} e_p
\end{displaymath}
and zero free eccentricity.
The coefficient $c'$ sets the secular precession rate.

We can perform canonical transformations to new variables
\begin{eqnarray}
x &=& \sqrt{2 \Gamma} \cos\varpi  - \sqrt{2 \Gamma_f} \cos\varpi_p \nonumber \\
y &=& \sqrt{2 \Gamma} \sin\varpi  - \sqrt{2 \Gamma_f} \sin\varpi_p \nonumber \\
I &=&  {x^2 + y^2 \over 2} \nonumber \\
\theta &=& \tan^{-1}(y/x).
\end{eqnarray}
These variables were also used by \citet{murray97}.
Here the momentum variable $I$ is related
to the particle's free or proper eccentricity rather than
the particle's eccentricity. 

Recovering the Hamiltonian (equation \ref{eqn:hamgen}) in the new variables
\begin{eqnarray}
K(\Lambda, \psi; I,\theta)
   &  = &
   a' \Lambda^2 + b'\Lambda +  c' I 
  \qquad
\label{eqn:hamgenI}
\\ 
& &  
 +  ~ {\rm constant } 
\nonumber
\\
& &  
  + ~ g_0 I^{1/2}   \cos{( \psi -  \theta })
\nonumber
\\
& &  
  + ~ (g_0 \Gamma^{1/2}_f + g_1 \Gamma_p^{1/2}) \cos{( \psi -  \varpi_p }).
\nonumber
\end{eqnarray}
The expansion about the fixed point associated with the forced
eccentricity does not change
the form of the term normally associated 
with the first order mean motion resonance, 
that  proportional to $\cos(\psi - \theta)$.   Consequently 
the resonance libration time and width are unchanged. 
The predicted semi-major axis where resonance overlap occurs
based on these resonance widths would be identical
to that for a planet in a circular orbit (see below).  

However the width of the corotation resonance differs
from that predicted using equation (\ref{eqn:hamgen}).
The coefficient describing the strength of this resonance 
($\propto  \cos{( \psi -  \varpi_p })$ in equation \ref{eqn:hamgenI}),
can be rewritten as 
\begin{displaymath}
\mu \sqrt{2} \alpha^{5/4} (f_{31} e_{forced}  + f_{27}e_p)
  \cos{( \psi -  \varpi_p }),
\end{displaymath}
using the coefficients listed in equations (\ref{eqn:gcofs}).
Since $f_{27}$ and $f_{31}$ have opposite signs 
the forced eccentricity term will tend to cancel the other term.
In the high $j$ limit the forced eccentricity equals the planet's
eccentricity and $f_{27} \approx - f_{31}$ so these two terms cancel.
The $\mu^{2/7}$ law is only valid for the high $j$ limit. Consequently
the corotation resonance completely cancels near the planet
for orbits with zero free eccentricity.
This implies
that the dynamics of particles at low
eccentricity near a planet in a circular orbit 
is similar to the dynamics of particles
at low free eccentricity near a planet in an eccentric orbit.

We note that our expansion above is valid only to first
order in the particle and planet eccentricity.  At high particle
and planet eccentricity, additional resonance terms become important,
and the low eccentricity expansion is no longer valid.
The above Hamiltonian (equation \ref{eqn:hamgenI}) 
is not restricted to zero free eccentricity but to 
low values of the free eccentricity due to the low degree of
the expansion.

\subsection{Rederiving the 2/7th law and the eccentricity 
dispersion in the disk edge}

In section 2 we numerically measured the eccentricity 
dispersion in the disk edge,
finding that it too does not significantly depend upon planet eccentricity.
Outside the chaos zone, planetesimals still experience perturbations from
the planet.
These perturbations have a characteristic size
set by size of perturbations in the nearest mean-motion
resonance that is not wide enough to overlap others and so is not part
of the chaotic zone.
Since particles in the edge reside outside
the chaotic zone, the velocity dispersion does not
increase with time.
In this subsection we check that our formulation can correctly 
predict the location of resonance overlap.    We then
predict eccentricity variations that would be predicted
by considering the role of the last resonance that
is not part of the chaotic zone.   

The width of the resonance can be thought of as the range of
semi-major axis over which the resonance has a large effect.
To estimate the first order resonance width  we rescale
the momentum and put the Hamiltonian
in a unitless form (e.g., as done by \citealt{M+D} in section 8.8 or
by \citealt{quillen06a} in section 3).
The factors used to rescale the Hamiltonian set the resonance width.
We perform a canonical transformation
of the Hamiltonian given in equation (\ref{eqn:hamgenI})
lacking the corotation term or 
\begin{eqnarray}
K(\Lambda, \psi; I,\theta)
   &  = &
   a' \Lambda^2    + b' \Lambda  + c' I 
  \qquad
\nonumber
\\ 
& &  
  + ~ g_0 I^{1/2}   \cos{( \psi -  \theta })
\nonumber
\end{eqnarray}
with generating function 
\begin{displaymath}
F_2  = J_1 (\theta - \psi ) + J_2 \psi
\end{displaymath}
leading to new variables 
\begin{eqnarray}
J_2- J_1 &=& \Lambda, \qquad \phi =\theta - \psi  \nonumber \\
J_1 &=& I, \qquad  \psi  = \psi \nonumber
\end{eqnarray}
and new Hamiltonian 
\begin{eqnarray}
K'(I,\phi;J_2,\psi) & = & a' \left(I^2  + J_2^2\right)
         + \left( c' - 2 a' J_2  - b'  \right) I
             \nonumber \\
&& + b' J_2 + g_0 I^{1/2} \cos \phi
\label{oneterm_g}
\end{eqnarray}
Note that $J_2$ is conserved and is small
for initial conditions near resonance with small initial
free eccentricity (or $I$).
Dropping constant terms and setting
\begin{equation}
B = c' - 2 a' J_2 - b',
\label{eqn:B}
\end{equation}
the Hamiltonian in Equation (\ref{oneterm_g}) becomes
\begin{displaymath}
K'(I,\phi)
     =  a' I^2 + B  I
       +  g_0   I^{k/2} \cos \phi.  
\end{displaymath}
Here the coefficient $B$ determines the distance from resonance.
By rescaling momentum and time
\begin{eqnarray}
\bar I  &=& \left|{ g_0 \over a' }\right|^{-2/3}
    I \nonumber \\
\tau &=&  |g_0|^{2/3}
             \left|a'\right|^{1/3}
                     t,
\label{eqn:rescale}
\end{eqnarray}
we can write this as
\begin{equation}
\bar{K}(\bar I, \phi)   = {\bar I}^2
        + \bar b \bar I   - {\bar I}^{1/2} \cos \phi
\label{eqnKp}
\end{equation}
where 
\begin{equation}
{\bar b} = B  |g_0|^{-2/3}
             \left|a'\right|^{-1/3}
\label{eqn:barb}
\end{equation}
sets the distance from exactly on resonance.
The resonance is only strong over a range 
$\Delta {\bar b} \sim 1$  (e.g., see Figure 8.10 by \citealt{M+D})
corresponding to a range of particle mean motions. 
Assuming slow secular precession and neglecting the term
$\propto J_2$,  the variation  $\Delta B \sim - \Delta b'$ 
(equation \ref{eqn:B}).
Equation \ref{eqn:coefs} allows us to relate $\Delta b'$
to the range of mean motions over which the resonance is
strong $\Delta b' \sim j \Delta n$. 
Equation \ref{eqn:barb} then implies that the
resonance is strong over a range of mean motions of size
\begin{displaymath}
\Delta n \sim j^{-1} \left|{ g_0  }\right|^{2/3}
               \left|{ a'  }\right|^{1/3}.
\end{displaymath}
For resonances near the planet we can use  
the asymptotic limit ($\alpha \to 1$, $j$ large) that gives
$g_0 \to \mu j$.  Subbing in for $g_0$ and considering
the range of semi-major axis rather than mean motion,
the resonance width is
\begin{displaymath}
\Delta a \sim 
   \mu^{2/3} j^{1/3} 
\end{displaymath}
where we have used $\Delta n \sim 3/2 \Delta a$.
Using a spacing between $j:j-1$ resonances of $\Delta a \sim {2\over 3} j^{-2}$ 
we find  that the resonances overlap at the resonance with
\begin{displaymath}
j \sim \mu^{-2/7}. 
\end{displaymath}
The semi-major axis corresponding to the $j:j-1$ resonance
at the chaotic zone boundary is set by the above $j$.
The $j:j-1$ mean motion resonance located at a semi-major axis of
$a = (1- 1/j)^{-2/3} \sim 1 + {2\over 3j}$ for large $j$.  
The distance between the planet and chaotic zone boundary, $\delta a_z$,
in semi-major axis, i.e., 
chaotic zone width, is then
\begin{equation}
\delta a_z \sim \mu^{2/7},
\end{equation}
recovering the 2/7 law \citep{wisdom80,duncan89,murray97,mudryk06}.

The eccentricity change or libration width
in the resonance just outside the boundary
would have size
\begin{equation}
I \sim \left|{ g_0  \over a'}\right|^{2/3}
\end{equation}
where we have used the libration width (given by the scaling
factor for $\bar I$ in Equation \ref{eqn:rescale}) 
or 
\begin{displaymath}
I \sim \mu^{2/3} j^{-2/3}.
\end{displaymath}
Subbing in $j \sim \mu^{-2/7}$ at 
the chaotic zone boundary gives $I \sim \mu^{6/7}$
and an eccentricity dispersion of
\begin{equation}
u_e \sim \mu^{3/7} 
\label{eqn:ue}
\end{equation}
just outside the chaos zone.
This dispersion could set the slope of the density distribution
in the disk edge (e.g,  \citealt{quillen06b}) .
The scaling predicted by Equation \ref{eqn:ue} is shown
compared to numerical measurements of the  eccentricity dispersion
in the disk edge in Figure \ref{fig:ue}.  
It provides a good fit to the measurements
and is independent of planet eccentricity as predicted by considering equation
\ref{eqn:hamgenI} for orbits with low free eccentricity.

\section{Summary and Discussion}

In this paper we have investigated the dynamics of low
free eccentricity collisionless
massless particles in the plane near a planet  on a eccentric orbit.
By determining the semi-major axis at which the particle lifetime
increases, we measure the width of the chaotic zone near
the planet.
For eccentricity $e_p < 0.3$ we find that the chaotic zone
width is independent of the planet's eccentricity and
matches that predicted by the 2/7 law.
The eccentricity dispersion in the disk edge and
the lifetime of particles within the chaotic zone is also nearly independent
of planet eccentricity.  These results suggest there is a similarity
in the dynamics of particles at zero free eccentricity 
and those at zero eccentricity
near a planet in a circular orbit. 

To account for our numerical results we have explored  the dynamics
of a  Hamiltonian system that takes into account
first order secular perturbations and the two terms that comprise
each first order mean motion resonance.  With a coordinate transformation
we have rewritten the Hamiltonian in terms of an action variable
that depends on the free eccentricity rather than 
the eccentricity.  At low free eccentricity
we find that the new Hamiltonian
resembles the Hamiltonian of a low eccentricity particle
near a planet in a circular orbit.
This accounts for the lack of sensitivity of the particle
dynamics on planet eccentricity.

For orbits with eccentricity equal to the forced eccentricity 
there is a  region in the plane with longer lived orbits  (compared to the
planet orbital period)
in close proximity to eccentric planets. Three dimensional simulations
that incorporate collisions are needed to see if these orbits tend
to be populated by long lived particles, as proposed
for Fomalhaut's eccentric ring \citep{quillen06b}.

\vskip 0.1truein
We thank the referee, Dr. Tsiganis, 
for numerous comments that have significantly 
improved this manuscript.
Support for this work was in part
provided by National Science Foundation grants AST-0406823 and PHY-0552695
the National Aeronautics and Space Administration
under Grant No.~NNG04GM12G issued through
the Origins of Solar Systems Program,
and HST-AR-10972 to the Space Telescope Science Institute.

{}

\end{document}